\documentclass[11pt,a4paper]{article}

\usepackage{graphicx}
\usepackage{amsmath,amsfonts,amsthm}

\catcode`@=11
\@addtoreset{equation}{section}

\catcode`@=12

\begin{document}
\pagenumbering{roman} \setcounter{page}{0} 
\title{Collapse transition of self-avoiding trails on the square lattice}
\author{A. L. Owczarek$^1$ and T. Prellberg$^2$\thanks{{\tt {\rm email:}
aleks@ms.unimelb.edu.au,t.prellberg@qmul.ac.uk}} \\
         $^1$ Department of Mathematics and Statistics,\\
         The University of Melbourne,\\
         Parkville, Victoria 3052, Australia.\\
$^2$School of Mathematical Sciences\\ 
Queen Mary, University of London\\
Mile End Road, London E1 4NS, UK
}
\date{
\begin{center}
        \today 
\end{center}
}

\maketitle 
 
\begin{abstract} 
  The collapse transition of an isolated polymer has been modelled by
  many different approaches, including lattice models based on
  self-avoiding walks and self-avoiding trails. In two dimensions, 
  previous simulations of kinetic growth trails, which
  map to a particular temperature of interacting self-avoiding trails,
  showed markedly different behaviour for what was argued to be the
  collapse transition than that which has been verified for models
  based of self-avoiding walks. On the other hand, it has been argued
  that kinetic growth trails represent a special simulation that
  does not give the correct picture of the standard equilibrium model.
  In this work we simulate the standard equilibrium interacting
  self-avoiding trail model on the square lattice up to lengths over
  $2,000,000$ steps and show that the results of the kinetic growth
  simulations are, in fact, entirely in accord with standard
  simulations of the temperature dependent model. In this way we
  verify that the collapse transition of interacting self-avoiding
  walks and trails are indeed in different universality classes in two
  dimensions.

\vspace{1cm} 
 

\end{abstract} 

\vfill

\newpage

\pagenumbering{arabic}

\section{Introduction} 
\setcounter{page}{1}

Over the past 25 years various lattice models of a single
self-interacting polymer chain have been analysed in both two and
three dimensions (and beyond). These include various types of
self-interacting self-avoiding walk, self-interacting trails
\cite{shapir1984a-a} (lattice paths that can intersect at a lattice
vertex but not along a lattice edge) and self-interacting random
(fully self-intersecting) walk models. The fundamental physical phase
transition \cite{gennes1979a-a} that these models are compared to is
that of the collapse of single polymer in a poor solvent as the
temperature is lowered. The question that arises when considering the
bulk of these studies is how robust is the universality class of the
collapse transition. The standard theory
\cite{gennes1975a-a,stephen1975a-a,duplantier1982a-a} of the collapse
transition is based on the $n\rightarrow 0$ limit of the magnetic
tricritical $\phi^4-\phi^6$ O($n$) field theory and related Edwards
model with two and three body forces
\cite{duplantier1986b-a,duplantier1987d-a}, which predicts an upper
critical dimension of three with subtle scaling behaviour in that
dimension. As an analysis of these theories is not exact in two
dimensions, lattice models form the basis of our knowledge. Analyses of
both two- and three-dimensional self-interacting trails
\cite{owczarek1995a-:a,prellberg1995b-:a} indicate that
they are in a different universality class to that of self-interacting
self-avoiding walks in those respective dimensions. There is no clear
understanding of why this is the case if true. A complication of this
scenario occurs because the numerically most extensive work on trails
\cite{owczarek1995a-:a,prellberg1995b-:a}, and that which draws the
conclusion of separate universality classes, uses so-called ``kinetic
growth'' simulations, or ``smart kinetic trails'', to study the
collapse point of self-interacting trails.  These simulations are of
trails produced in such a way that one argues they form a distribution
of self-interacting trails at one particular temperature. One then
further argues from the numerical evidence that this temperature is
precisely the collapse temperature. Now, it was claimed
\cite{grassberger1996a-a} that the collapse transition associated with
``smart kinetic trails'' is first-order, though no evidence of this
could be found in two dimensions. The explanation of this relates to
the vanishing of renormalised three-body interactions in the smart
kinetic trails. It was further suggested in \cite{grassberger1996a-a}
that studying the smart kinetic trails was misleading when considering 
the full equilibrium self-interacting trail model.

To try to see if these arguments hold in two dimensions we have
simulated self-avoiding trails over a range of temperatures near the
collapse point. The collapse temperature is indeed the point onto
which the smart kinetic trails map, and we compare these with our
earlier results concerning two-dimensional trail collapse based on the smart
kinetic trail simulations.

\section{The Model}

The model of self-interacting trails (ISAT) on the square lattice is
defined as follows. Consider all different bond-avoiding paths
$\varphi_N$ of length $N$ that can be formed on the square lattice
with one end fixed at a particular site (the set $\Omega_N$). Associate
an energy $-\varepsilon$ with each doubly-visited site. For each
configuration $\varphi_N$ count the number $m(\varphi_N)$ of
doubly-visited sites of the lattice and give that configuration a
Boltzmann weight $\omega^m$, where $\omega=\exp(\beta\varepsilon)$. The
partition function of the ISAT model is then given by
\begin{equation}
Z_N(\omega) = \sum_{\varphi_N \in \Omega_N} \omega^{m(\varphi_N)} \; .
\end{equation}
The reduced and normalized internal energy
\begin{equation}
U_N = \frac1N\langle m \rangle
\end{equation} and reduced and normalized
specific heat 
\begin{equation}
C_N=\frac1N\left( \langle m^2 \rangle - \langle m \rangle^2\right)
\end{equation} are defined in the
usual way. 

It was argued in \cite{meirovitch1989d-a,bradley1990a-a} that the
smart kinetic growth trails simulates at a specific Boltzmann weight
$\omega=\omega_c\equiv 3$ and it was later argued
\cite{owczarek1995a-:a} that $\omega_c$ is indeed the collapse
value of the Boltzmann weight.

Let us assume that the phase transition occurring at $\omega_c$ is
critical and furthermore that the specific heat diverges at the
transition (all previous work supports this second proposition). 
Now we define the exponent $\alpha$ from the divergence of the 
thermodynamic limit specific heat as usual as
\begin{equation}
c(\omega) = \lim_{N\rightarrow\infty} C_N \sim |\omega_c -\omega|^{-\alpha}
\end{equation}
Let us return to finite $N$ scaling and consider the peak in the
specific heat near $\omega_c$. Let the peak be located at
$\omega_{c,N}$ and define the width of the transition $\Delta\omega$
by the difference in the omega values at the half-heights (that is,
the values of $\omega$ that give values of $C_N$ as
$C_N(\omega_{c,N})/2$). The crossover exponent is defined by 
\begin{equation}
\Delta\omega \sim  \frac{a}{N^\phi}
\end{equation}
where $a$ is a constant, and, in turn,  gives us a scaling variable $(\omega
-\omega_c)N^\phi$ on which to scale for data collapse. The standard
scaling theory \cite{brak1993a-:a} predicts that the crossover
exponent $\phi$ is related to the specific heat exponent $\alpha$ via
the scaling relation
\begin{equation}
\label{scalerel}
2 -\alpha=1/\phi
\end{equation}
and that the shift of the peak from its thermodynamic limit value is
also governed by $\phi$, that is
\begin{equation}
|\omega_c-\omega_{c,N}| \sim \frac{b}{N^\phi}
\end{equation}
where $b$ is a constant. Moreover, the peak value of the finite size specific heat
should then behave as \cite{brak1993a-:a}
\begin{equation}
 C_N(\omega_{c,N}) \sim dN^{\alpha \phi} \; ,
\end{equation}
where $d$ is a constant.  Using all of the above one can obtain local estimates of
the exponents $\alpha$ and $\phi$ from this form, and these can be
independently verified by studying the width and the shift of the
transition via the specific heat peak.

A further independent test is to study the internal energy as it is
predicted to scale as 
\begin{equation}
U_N(\omega_{c,N})\sim U_\infty(\omega_c) - e N^{(\alpha-1)\phi} \; ,
\end{equation}
where $U_\infty$ is the thermodynamic limit internal energy and $e$
is a constant. Usually this would be more difficult to analyse as
$U_\infty(\omega_c)$ is unknown but for this model we can argue that
$U_\infty=0.4$ as follows from a kinetic growth argument.

Consider an $N$-step loop which occupies $M$ lattice sites. The 
number of contacts is given by the number $m$ of doubly visited sites,
and we have $m=N-M$. Any site of this loop could have been the starting
point, and in order for a site to be visited twice, the loop must not 
have been closed at the first return visit. The probability of not closing 
upon the first return is $2/3$, so that for large loops $m/M\rightarrow2/3$, from whence
it follows that $m/N\rightarrow2/5$.

In our previous work \cite{owczarek1995a-:a} we estimated
\begin{equation}
\phi=0.88^{+0.07}_{-0.05}
\end{equation}
and so $\alpha\approx 0.86$. This is equivalent to the value of $
\alpha\phi \approx 0.76$
for the exponent describing the divergence of the specific heat
peak. We note that the established values for self-interacting
self-avoiding walks in two dimensions \cite{duplantier1987a-a}, that
is, $\phi=3/7$ and $\alpha=-1/3$,  imply that the specific heat does not diverge!
On the other hand, if there was a first order transition $\alpha=\phi=1$.

\section{Results}
We have used the now standard PERM algorithm \cite{grassberger1997a-a}
to simulate ISAT at various fixed temperatures around $\omega_c=3$ for
trail lengths from $2^{15}=32768$ (denoted as $32$K) in factors of $4$
up to $2^{21}= 2097152$ (denoted as $2048$K). We chose the range of
temperatures by first simulating at $\omega=3$ and then reweighting
the obtained histogram
to give an estimate of the location of the specific heat peak $\omega_{c,N}$.
We then repeated the simulation at $\omega_{c,N}$. The multi-histogram method 
\cite{ferrenberg1988a-a} was
then used to give data throughout the transition region. The
simulations at $\omega=3$ gave almost identical results 
to those additionally using the peak data. The
quantities of interest such as $\Delta\omega$ and $\omega_{c,N}$ were re-estimated.

\begin{figure}
\begin{center}
\includegraphics[width=12cm]{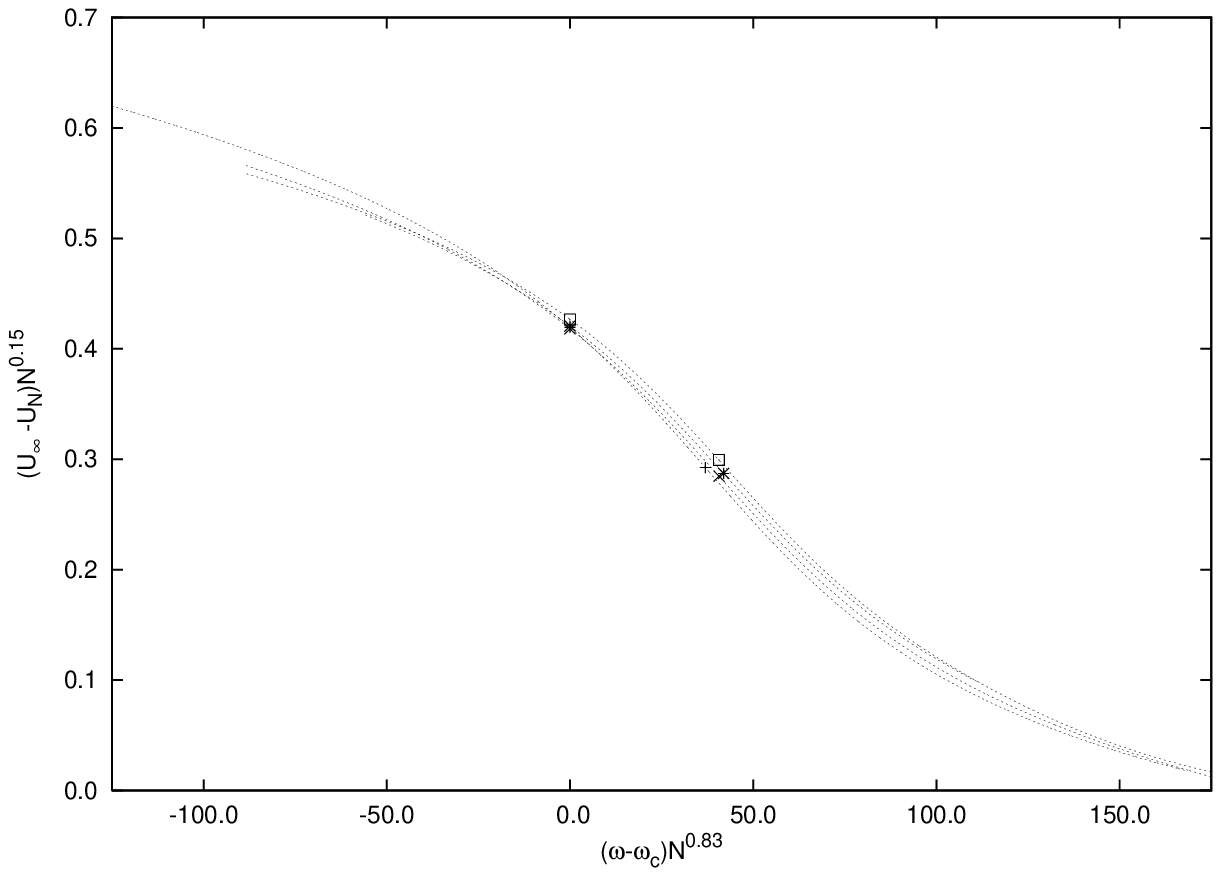}
\end{center}
\begin{center}
\includegraphics[width=12cm]{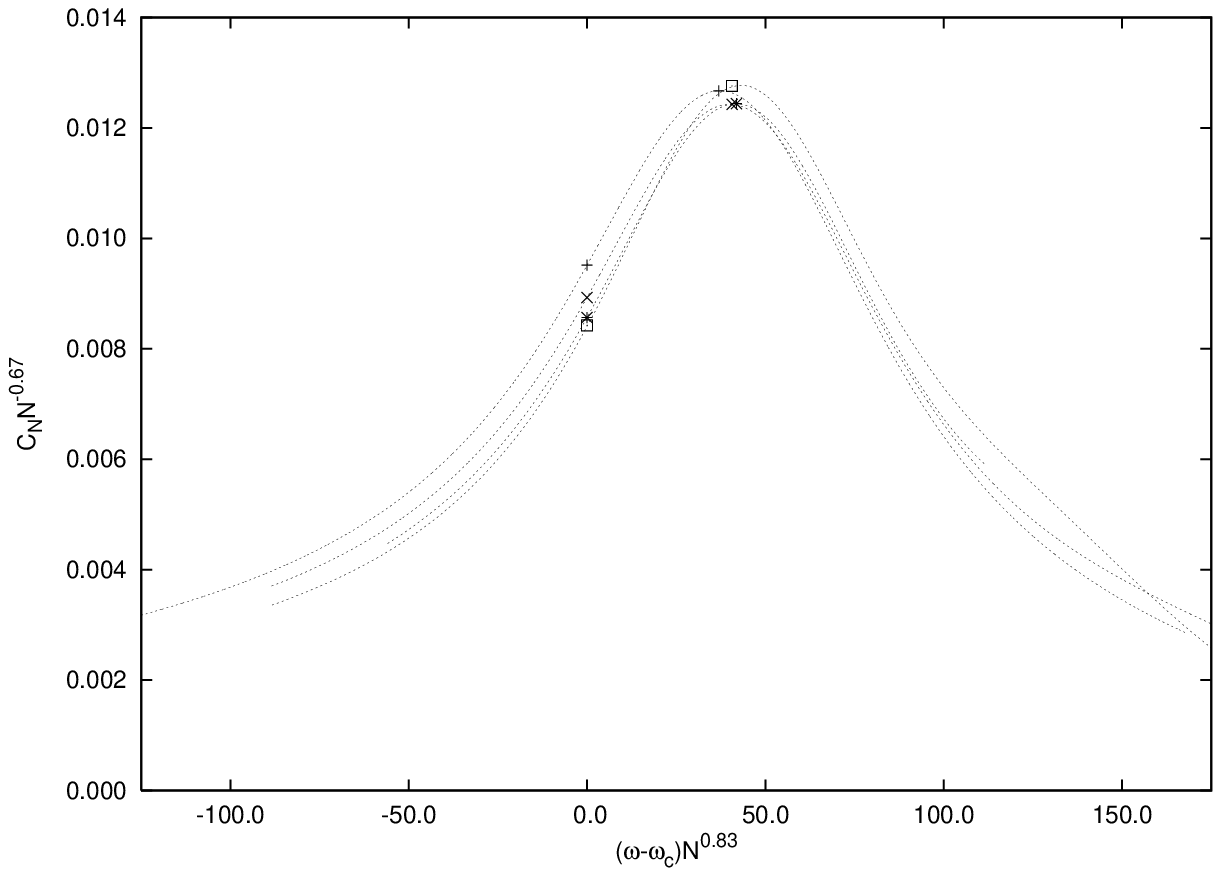}
\end{center}
\caption{\it Scaled internal energy gap $(U_\infty -
  U_N)N^{(1-\alpha)\phi}$ and specific heat $C_N N^{-\alpha\phi}$ versus scaled $(\omega-\omega_c)N^\phi$ 
  for lengths $32K$, $128K$, $512K$, and $2048K$, comparing the
  individual data points with the results from the multi-histogram
  method.}
\label{figure1} 
\end{figure}

We attempted to scale all the data using a consistent set of
exponents. The best fit for the specific heat data was obtained for 
\begin{equation}
\alpha\phi=0.6\dot{6} \quad \mbox{ so } 
\alpha= 0.80 \mbox{ and } \phi= 0.8\dot{3} 
\end{equation}
while for the internal energy data the best fit used
\begin{equation} 
\phi = 0.85\quad \mbox{ so } \alpha \approx 0.82\mbox{ and }\alpha\phi=0.7   
\end{equation} 
Clearly there is an error of at least 0.025 in these estimates and the
statistical spread confirmed an error of about 0.03.
We have used the scaling relation (\ref{scalerel}) for
consistency.  The first figures show the the scaling of various
quantities using those exponent assumptions. In figure (\ref{figure1})
the scaled difference of the internal energy to its thermodynamic
value $(U_\infty - U_N)N^{(1-\alpha)\phi}$ and scaled specific heat
$C_N N^{-\alpha\phi}$ are plotted against the scaling variable
$(\omega-\omega_c)N^\phi$. These demonstrate consistency in three ways:
between the specific heat, the internal energy and the horizontal
scaling variable all at once.

\begin{figure}
\begin{center}
\includegraphics[width=12cm]{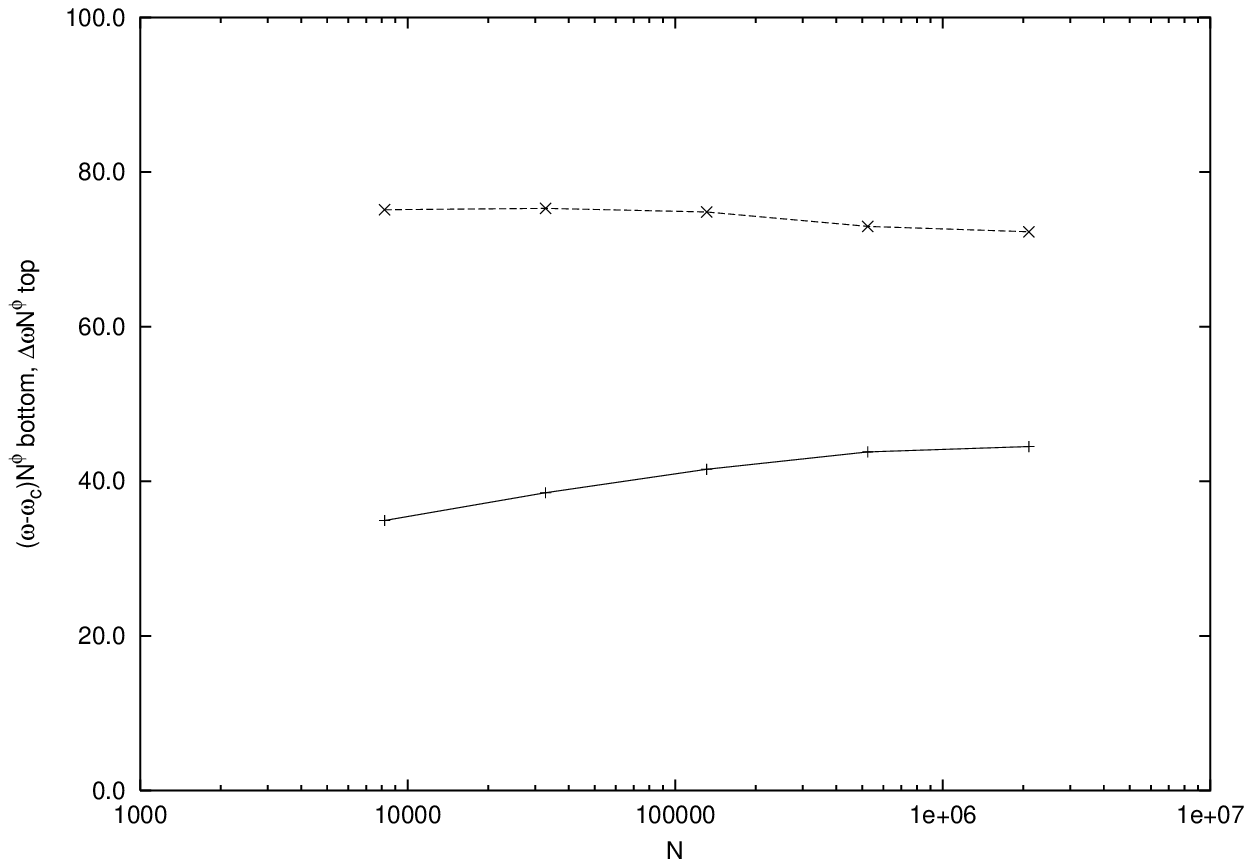}
\end{center}
\caption{\it Scaling of the transition: shift and width of the
  collapse region. Shown are the scaling combinations
  $N^{\phi}(\omega_{c,N}-\omega_c)$ and $N^{\phi}\Delta\omega$ versus
  $N$.  We have that $\omega_c=3$ and have used the estimate of
  $\phi=0.84$.}
\label{figure2} 
\end{figure}

We have estimated the crossover exponent independently via both the
shift $(\omega_{c,N}-\omega_c) $ and the width $\Delta\omega$ of the
transition. Figure (\ref{figure2}) shows plots of the appropriate
scaling combinations for these two quantities against $N$ using the
estimate $\phi=0.84$ (in between our previous estimates), being the best value
using these plots.

\begin{figure}
\begin{center}
\includegraphics[width=12cm]{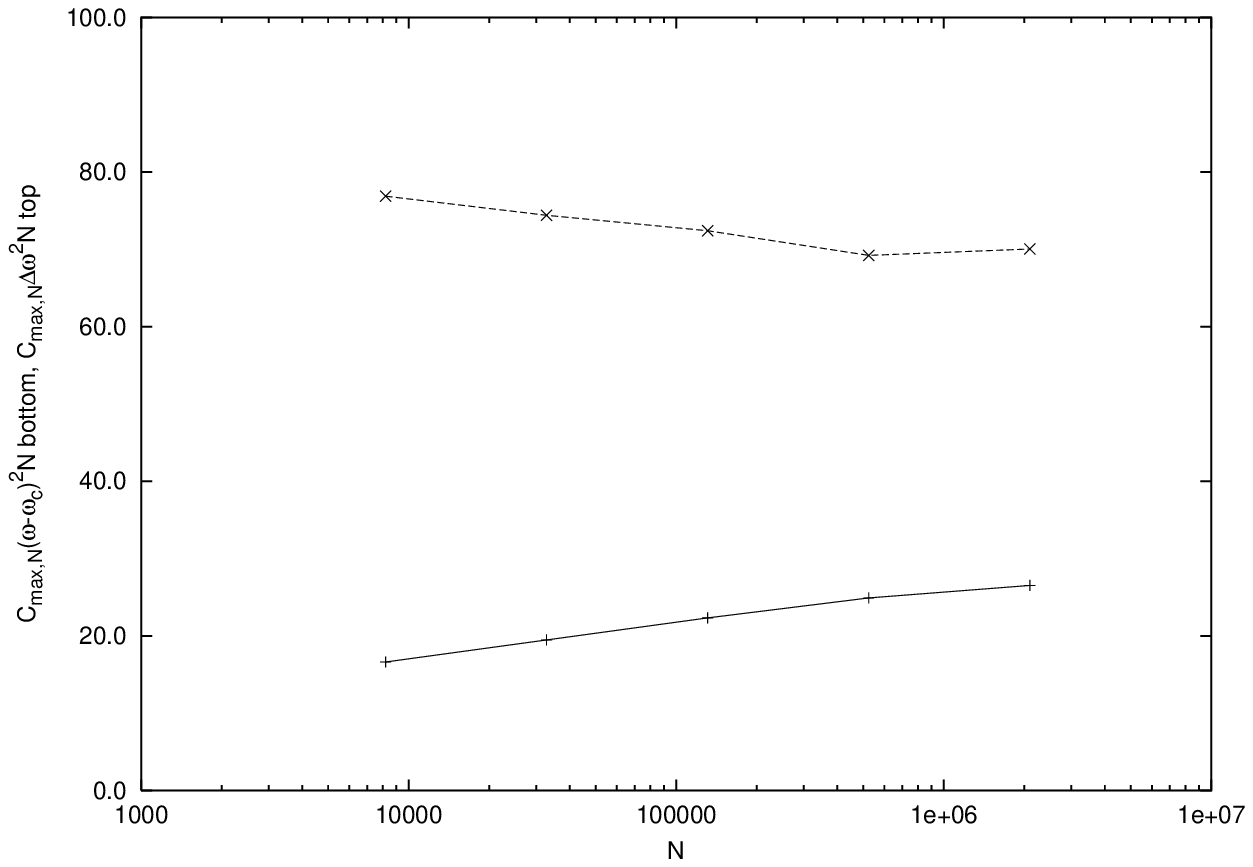}
\end{center}
\caption{\it Scaling of the transition: height of the specific heat
  peak. Shown are the scaling combinations
  $C_N(\omega_{c,N}-\omega_c)^2$ and $C_N\Delta\omega^2$ versus $N$.
  We have that $\omega_c=3$.}
\label{figure3} 
\end{figure}

Finally, in Figure (\ref{figure3}) we demonstrate the convergence of
the scaling by plotting scaling combinations that should be constant
in the absence of corrections-to-scaling. These give us confidence in
the results from the other figures.

We therefore conclude that our new simulations away from the ``smart
kinetic growth point'' are in agreement with the exponent estimates of
the smart kinetic growth simulations with perhaps slightly smaller
values of $\phi$ and $\alpha$ as previous central estimates. We
now estimate
\begin{equation} 
\alpha= 0.81(3) \quad \mbox{ and } \phi= 0.84(3)
\end{equation}

\section{Conclusion}

We have simulated self-interacting self-avoiding trails on the square
lattice up to lengths of $2,097,152$ for a range of temperatures around
the collapse transition temperature. We conclude that the results are
in good agreement with earlier simulations based on smart kinetic
trails. They demonstrate a phase transition which is neither of the
type displayed by self-interacting self-avoiding walks nor is it first
order, although it is very strong, as predicted by an analysis of the
smart kinetic growth trails.

\section*{Acknowledgments} 
 Financial support from the Australian Research Council and the Centre
 of Excellence for Mathematics and Statistics of Complex Systems is
 gratefully acknowledged by the authors.

\end{document}